# IS THERE A THIRD COMPONENT IN THE INTERMEDIATE POLAR V405 AUR?

Vitalii V. Breus[1], Ivan L. Andronov[1], P. Dubovsky[2], S. V. Kolesnikov[3],
E. A. Zhuzhulina[4], T. Hegedus[5], P. Beringer[5], K. Petrik[6],
J. W. Robertson[7], A.V. Ryabov[3], V.G. Tsehmeystrenko[8], I. Kudzej[2],
N. M. Shakhovskoy[9]

[1] Department "High and Applied Mathematics", Odessa National Maritime University
[2] Vihorlat Astronomical Observatory, Humenne, Slovakia
[3] Astronomical Observatory, Odessa National University, Odessa, Ukraine
[4] Department of Astronomy, Odessa National University, Odessa, Ukraine
[5] Baja Astronomical Observatory, Baja, Hungary
[6] Astronomical Observatory, Hlohovec, Slovakia
[7] Arkansas Tech University, Russellville, USA
[8] Amateur Astronomical Observatory "Heavenly Owl", Odessa, Ukraine, http://www.**heavenly-owl**.com.ua
[9] Crimean Astrophysical Observatory, Nauchny, Ukraine

**Abstract**

Variability of the spin period of the white dwarf in the V405 Aur (RX J0558.0+5353) system using our observations and previously published maxima timings is analyzed. As the phase light curve contains two nearly equal photometric waves, one maximum was set as a "primary" one. The ephemeris for the maxima of the "spin variability" (due to rotation of the magnetized white dwarf) for recent seasons 2010-2012 is $T_{max}$ = HJD 2455882.470614(25)+0.00631314602(46)·$E$. This corresponds to a significant negative trend at the "O-C" diagram. Due to significant gaps in the observational data and statistical error of timings, there may be some suggestions on the spin period variability – a fast period "jump" in 2007y; secular period variations; a cubic ephemeris (which may be interpreted by a precession of the magnetic white dwarf at a time-scale of decades) or a periodic change with a period of 6.2 years and semi-amplitude of 17.2±1.8 sec. For the present observations, more reliable are two latter models. To distinguish between them, a continuation of monitoring is needed. The periodic variations may be interpreted by a light time effect caused by a third body of mass ($M_3 \geq 0.09 M_\odot$), which corresponds to a low-mass star, but not to an extra-solar planet. In this case, the system belongs to a rare class of cataclysmic variables with a third body.

**Introduction**

The intermediate polar V405 Aur was discovered as an optical counterpart of the soft ROSAT source 1RXS_J055800.7+535358 by Haberl et al. [1]. The soft X-Ray flux was changing with a period of 272.74s, which was supposed to be a spin period of the white dwarf. The presence of optical pulsations at a period of 272.785s was reported by Ashoka et al. [2].

Later Allan et al. [3] and Skillman [4] made independent announcements that the spin period of the white dwarf is twice longer (545.45s).

The double value of the spin period was justified by detection of circular polarization with a period of 544.4s (Shakhovskoj and Kolesnikov [5]). Noting very similar photometric maxima (two during a polarimetric period), Shakhovskoj et al. [6] suggested a nearly-equatorial location of two accretion columns.

Evans and Hellier [7] discussed the double-peaked spin pulse. The shape of the phase curve is wavelength-dependent, with a strong separation between the peaks at soft X-Ray range and a "saw-tooth" shape at hard X-Rays. This is naturally explained by an inequality of the columns.

Piirola et al. [8] analyzed the maxima timings obtained in 1994-2007 and published second-order polynomial fit to the maxima timings

$$T_E = 2449681.46389(5) + 0.0063131474(4)E + 4(4)*10^{16}E^2.$$

Since the second-order coefficient was not statistically significant, they also published a linear ephemeris:

$$T_E = 2449681.46387(2) + 0.0063131476(3)E$$

**Observations**

We obtained photometric CCD observations using different telescopes: 1m VNT (filters V and R) and 28cm Pupava (unfiltered) reflectors in Vihorlat Astronomical Observatory, Humenné, Slovakia, 35cm BAT and 50cm reflector in Baja Astronomical Observatory, Hungary (V and R filters) and 20cm MEADE LX-200 at the Observatory and Planetarium in Hlohovec, Slovakia (V and R filters). Also we used UBVRI photometry obtained at the 1.25m AZT-11 and wide-R photometry at the 2.6m Shain Telescope at the Crimean Astrophysical Observatory, Nauchny, Ukraine and unfiltered observations from the Arkansas Tech University Observatory. Recently, the "Heavenly Owl" amateur astronomical observatory was put into operation [9], where the monitoring was carried out using the V filter. The final time series were obtained using the program MCV (I.L.Andronov, A.V.Baklanov [10]) taking into account multiple comparison stars. Periodogram analysis was carried out using the programs MCV and FDCN (Andronov [11]).

**O-C Analysis**

For phase curves, we have used a preliminary value of the spin period of $P = 0.^d0063131474$ and initial epoch 2449681.46389 (Eq. (1) of [8]). Using this period, we have used a second-order trigonometric polynomial fit to the magnitudes. This corresponds to a double photometric wave per one polarimetric one [5,6]. As both photometric waves appear to be nearly identical, we suggested to use the timing (or corresponding phase) of the brightness maximum (minimum of stellar magnitude) of the harmonic of the main period (one per night) instead of timings of the individual maxima, the accuracy of which is much worse than that of the "nightly" timing. Anyway, at the figures one may see a drastic scatter of phases, especially for the SuperWASP data (www.superwasp.org).

Contrary to a classical representation of the "O-C diagram" as a dependence of the timings from an ephemeris on the cycle number $E$, we have used phases instead, i.e. $\phi = (O - C)/P$. For a correct ephemeris, the phases should be concentrated near the zero value. In our case, the times of maxima estimated by a program correspond to a harmonic with a double frequency, thus it is expected that phases like -0.5, +0.5 correspond to the same "zero", and thus one may make corrections for these values. For a complete set of "our+published" data, the phases seem to range from -0.13 to 0.11.

Contrary to a suggestion of Piirola et al. [8], the points for the recent years show a distinct period decrease. A simplest hypothesis is that the period has underwent change (approximately in 2007 ($E \approx 713491$)). We analyzed separately the photometric data, obtained at the Vihorlat Astronomical Observatory in 2010-2013. As light curves in V and R filters had comparable amplitudes but different stellar magnitudes, we subtracted the mean value of stellar magnitude from the data in each filter and joined it. We determined the new value of period and initial epoch that better corresponds to all spin maxima timings in our observations.

$$T_{max} = HJD\ 2455882.470614(25) + 0.00631314602(46) \cdot E$$

However, previous studies of intermediate polars argue for smooth period variations rather than period jumps. From theoretical expectations, the spin periods of the white dwarf should be equal to some equilibrium value, which is equal to period of "Kepler" rotation of the inner accretion disk at a distance of the magnetospheric radius ([13], [14]). Period variations may be caused by changes of the accretion rate due to modulation of the mass transfer caused by magnetic activity of the red secondary ([15]) fluctuations of the orbital separation [16], or precession of the magnetic white dwarf (which will be present either with constant, or variable accretion rate) [17,18]. At time scales of decades, one may see only a part of the curve of cyclic variations. Thus apparently the "O-C" diagram may be not a "wave", but a square (for smaller time intervals) or cubic parabola (for larger intervals).

So we analyzed other models for the period variations. To decrease the error estimate of the initial epoch and the period, we computed the ephemeris for a different integer epoch $E_0$, which is close to a sample mean of the observational values. In our case, $E_0 = 504600$. A weighted fit to the phases of maxima $\phi$ leads to the following quadratic ephemeris:

$T_{max} = HJD\ 2452867.07807(2) + 0.006313147426(70) \cdot (E - E_0) - 659(233) \cdot 10^{-18}(E - E_0)^2$.

The value of the quadratic term $Q$ reaches $2.8\sigma$, so its deviation from zero is not statistically significant. It doesn't fit our recent observations enough good, so we calculated the 3-rd order weighted fit to the phases of maxima and got the following cubic ephemeris:

$T_{max} = HJD\ 2452867.07807(2) + 0.006313147760(131) \cdot (E - E_0) - 502(237) \cdot 10^{-18}(E - E_0)^2$
$-239(80) \cdot 10^{-23}(E - E_0)^3$.

It corresponds to all observations better and fits most recent observations showing a distinct negative trend during recent years.

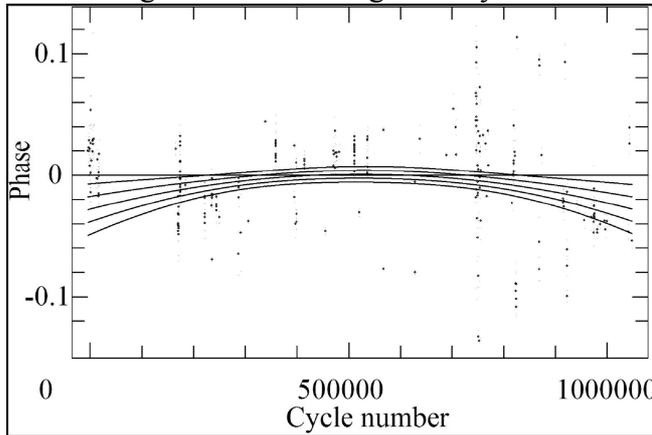

Figure 3: Dependence of phases of maxima timings on cycle number of the spin period: circles - original observations, line - an approximation using 2-nd order polynomial fit with corresponding $\pm 1\sigma$ and $\pm 2\sigma$ error corridors.

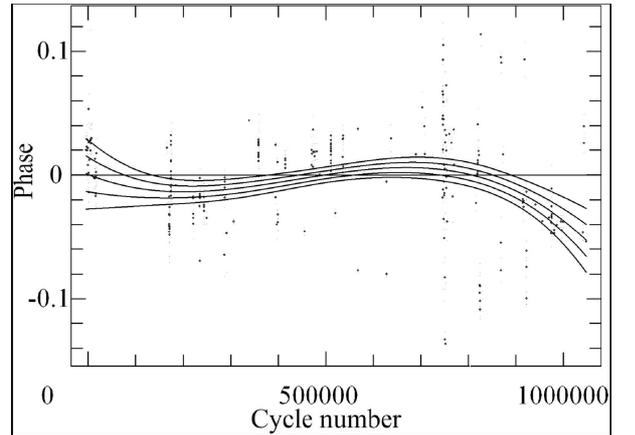

Figure 4: Dependence of phases of maxima timings on cycle number of the spin period: circles - original observations, line - an approximation using 3-rd order polynomial fit with corresponding $\pm 1\sigma$ and $\pm 2\sigma$ error corridors.

**Third body hypothesis**

An alternate model is a presence of a third body - a star or a massive planet. In this case, the theoretical "O-C" diagram is a periodic wave ("mono-harmonic" for a circular orbit or "multi-harmonic" for an elliptic orbit).

We calculated the periodogram using the approximation combining a 1-st order trigonometric and a 1-st order algebraic polynomials, using the program MCV. The maximum peak at the periodogram corresponds to a period of $2268^d = 6.2$yr. The corresponding fit is

$\phi = -0.00049(219) + 0.0000002(14) \cdot (T - 2452881) +$
$+ 0.0315(32) \cos(2\pi \cdot (T - 2452389)/2268)$

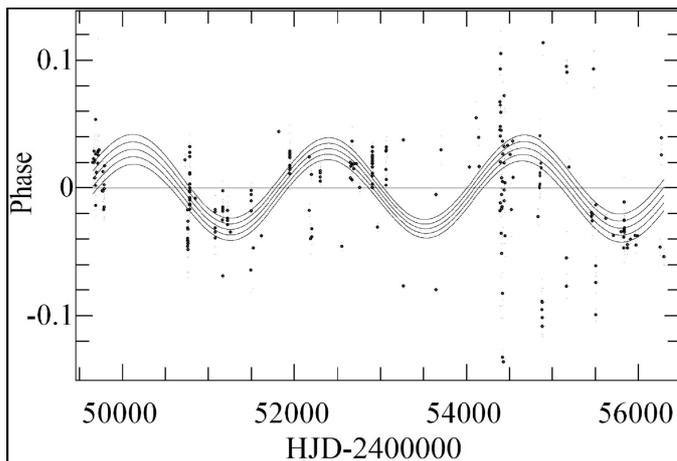

Figure 5: Dependence of phases of maxima timings on HJD: circles - original observations, line - an approximation using 1-st order trigonometric and algebraic polynomial fits with corresponding ±1σ and ±2σ error corridors.

These periodic variations are statistically significant (at a level of semi-amplitude of 9.7σ). However, the amplitudes of harmonics of this period are not statistically significant.

So, one may suggest a third body orbiting the inner binary system on a nearly circular orbit with a period of ≈ 6.2yr, with a distance of the center of masses to the binary of 17.2 ± 1.8 light seconds, or $(5.15 ± 0.53) \cdot 10^9$ meters). The corresponding mass function [12] is $f(M) ≈ 0.09 M_\odot$. As this is a lower limit of the mass of the third body, we may suggest that this is not a planet, but probably a red dwarf.

Contrary to the third-order polynomial, the sinusoidal fit shows a return to zero of phases at the end of the interval of observations. This difference will increase in future. Thus there may be a check-out, which model is correct, during even a year or two of subsequent observations.

Acknowledgements. The study is a part of the "Inter-Longitude Astronomy" campaign [19] and "Ukrainian Virtual Observatory" project [20]. The SuperWASP data have been analyzed in an addition to ours. VVB and ILA thank Dr. Bogdan Wszołek, the Institute of Physics of the Jan Długosz Academy in Częstochowa and the "Astronomia Nova" society for hospitality.